\documentclass[apj]{emulateapj}

\usepackage{apjfonts}

\usepackage{graphicx}
\usepackage{color}
\usepackage{amsbsy}
\usepackage{mathrsfs}
\usepackage{mathpazo,bm}
\usepackage{amsmath}
\usepackage{bigdelim}
\usepackage{bigstrut}
\usepackage{subfigure}
\usepackage{multirow}

\newlength{\hfwidth}
\newlength{\hfwidthsingle}
\newlength{\figspace}
\newcommand{\beq}{\begin{equation}}
\newcommand{\eeq}{\end{equation}}
\newcommand{\pderiv}[2]{\frac{\partial{#1}}{\partial{#2}}}
\newcommand{\deriv}[2]{\frac{\mathrm{d}{#1}}{\mathrm{d}{#2}}}

\newcommand{\aderiv}[1]{\frac{\mathrm{D}{#1}}{\mathrm{D}t}}

\newcommand{\Ma}{\mathrm{Ma}}

\newcommand{\ttimes}[1]{10^{#1}}
\newcommand{\xtimes}[2]{#1\times{10^{#2}}}

\renewcommand{\v}[1]{{\boldsymbol{#1}}} 

\newcommand{\del}{\v{\nabla}}
\newcommand{\grad}{\del}
\newcommand{\Div}{\del\cdot}

\newcommand{\Msun}{M_\odot}

\newcommand{\cv}{c_{_{V}}}
\newcommand{\taut}{\tau}
\newcommand{\cs}{c_\mathrm{s}}
\newcommand{\csO}{c_\mathrm{s0}}
\newcommand{\Mjup}{\,M_\mathrm{J}}  
\newcommand{\alphat}{\beta}
\newcommand{\tauref}{\tau_0}

\newcommand{\Eq}[1]{Eq.~(\ref{#1})}

\newcommand{\eq}[1]{\Eq{#1}}

\newcommand{\Fig}[1]{Fig.~\ref{#1}}
\newcommand{\fig}[1]{\Fig{#1}}

\newcommand{\Figss}[2]{Figs.~\ref{#1}--\ref{#2}}

\newcommand{\sect}[1]{Sect.~\ref{#1}}

\definecolor{brown}{rgb}{0.42,0.24,0.07}
\definecolor{darkgreen}{rgb}{0.0,0.6,0.00}
\definecolor{purple}{rgb}{0.7,0.0,0.7}
\definecolor{black}{rgb}{0.0,0.0,0.0}

\def\black#1{\textcolor{black}{#1}}

\newcommand{\acb}[1]{\black{{#1}}} 
\newcommand{\wlad}[1]{\black{{#1}}} 
\newcommand{\mordecai}[1]{\black{{#1}}} 
\newcommand{\alex}[1]{\black{{#1}}} 

\shorttitle{Buoyancy-unstable planetary wake}
\shortauthors{Richert et al.}

\begin{document}

\title{On shocks driven by high-mass planets in radiatively inefficient disks. I. \\
Two-dimensional global disk simulations}

\author{Alexander J.W. Richert\altaffilmark{1,2,3,4}, Wladimir Lyra\altaffilmark{3,4,5}, Aaron Boley\altaffilmark{6}
\\ Mordecai-Mark Mac Low\altaffilmark{7} \& Neal Turner\altaffilmark{3}}
\altaffiltext{1}{Department of Astronomy \& Astrophysics, Penn State University, 525 Davey Lab, University Park, PA 16802, USA. ajr327@psu.edu} 
\altaffiltext{2}{Center for Exoplanets \& Habitable Worlds, Pennsylvania State University.}
\altaffiltext{3}{Jet Propulsion Laboratory, California Institute of Technology, 4800 Oak Grove Drive, Pasadena, CA, 91109, USA. wlyra@jpl.nasa.gov, neal.j.turner@jpl.nasa.gov}
\altaffiltext{4}{Division of Geological \& Planetary Sciences, California Institute of Technology, 1200 E California Blvd MC 150-21, Pasadena, CA 91125 USA}
\altaffiltext{5}{Sagan Fellow}
\altaffiltext{6}{Department of Physics and Astronomy, University of British Columbia, 6224 Agricultural Road, Vancouver, BC V6T 1Z1, Canada. acboley@phas.ubc.ca}
\altaffiltext{7}{Department of Astrophysics, American Museum of Natural History, Central Park West at 79th Street, New York, NY 10024-5192, USA. mordecai@amnh.org}


\begin{abstract}
  Recent observations of gaps and non-axisymmetric features in the dust
  distributions of \wlad{transition} disks have been interpreted as
  evidence of embedded massive protoplanets. \wlad{However, \mordecai{comparing} the 
  predictions of planet-disk interaction \mordecai{models} to the observed features
  \mordecai{has shown}
  far from perfect agreement. This may be due to \mordecai{the strong}
  approximations \mordecai{used for the prediction}s.
  \acb{For example, spiral arm fitting typically uses results that are
    based on low-mass planets in an isothermal gas.}}  
  In this work, we \mordecai{describe} two-dimensional, global, hydrodynamical simulations
  of disks with embedded protoplanets\acb{, with and without the assumption
  of local isothermality, for} a range of planet-to-star mass
  ratios 1--10~$\Mjup$ for a 1~$\Msun$ star. We use the {\sc Pencil Code} in polar coordinates for
  our models.
  We find that the \mordecai{inner and outer spiral wakes} of massive
  protoplanets ($M\gtrsim5\,\Mjup$) \mordecai{produce significant shock
    heating that can trigger buoyant instabilities.  These drive}
  sustained turbulence throughout the disk when they occur. 
  \mordecai{The strength of this} effect \mordecai{depends} strongly on the mass of the planet and the
  thermal relaxation timescale; for a 10~$\Mjup$ planet
  embedded in a \wlad{thin,} purely adiabatic disk, the \wlad{spirals,} gaps\wlad{,} and vortices typically
  associated with planet-disk interactions are disrupted. We
  find that the effect is only weakly dependent on the initial radial
  temperature profile. \wlad{The spirals \alex{that} \mordecai{form in
      disks heated by the effects we have described may fit the 
      spiral structures observed in transition disks better than the spirals predicted} 
 \acb{by linear isothermal theory}.}
\end{abstract}
\subjectheadings{hydrodynamics --- planet-disk interactions ---
  planets and satellites: formation --- protoplanetary disks --- shock
  waves --- turbulence}

\section{Introduction}
\label{sect:introduction}

Decades of analytical and numerical work
\citep{Papaloizou-Lin84,Lin-Papaloizou86a,
  Lin-Papaloizou86b,Nelson00,Masset-Snellgrove01,Paardekooper-Mellema04,Quillen04,
  deValBorro06,Klahr-Kley06,Lyra09a,Zhu11,Kley12,Kley-Nelson12} have
established that massive ($\gtrsim1\Mjup$) protoplanets generate
observable gaps and other features in their host disks that can in
principle be resolved by ALMA \citep[see, e.g., the review
of][]{wolfetal2012}. These \mordecai{models} show that the
gravitationally-induced tides of protoplanets with masses
\mordecai{exceeding} Saturn\mordecai{'s} clear axisymmetric gaps in the
disk.  \mordecai{These} generate long-lived, non-axisymmetric vortices
due to \acb{the} Rossby wave instability (RWI) at the gap edges
\citep{Lovelace-Hohlfeld78,Toomre81,Papaloizou-Pringle84,Papaloizou-Pringle85,Hawley87,Lovelace99}.
Non-axisymmetric dust clouds have been observed in \wlad{transition}
disks \wlad{ \mordecai{at} sub-mm wavelengths}
\citep{Oppenheimer08,Brown09,Casassus12,Isella13,Casassus13,vandermareletal2013},
features that are usually interpreted as \wlad{vortices induced by}
embedded protoplanets.  \wlad{Spiral structures, one of the hallmarks
  of planet-disk \acb{interactions} (Goldreich \& Tremaine 1979,
  Ogilvie \& Lubow 2002, Rafikov 2002), are also seen in several
  transition \acb{disks} in polarized scattered light (Muto et
  al. 2012, Garufi et al. 2013, Avenhaus et al. 2014, Currie et
  al. 2014).}

However, because the \acb{putative corresponding} planets are not 
\mordecai{directly observed}, \acb{sufficiently detailed dynamical and
  radiative transfer models must be consistent with observations\mordecai{,} and
  all other possible ways to produce the observed features must be
  ruled out. } 
\acb{For example, there are several mechanisms that are known to generate 
vortices without a planet in the disk}. \acb{The} RWI \acb{can occur} at dead zone boundaries \citep{Varniere-Tagger06,lyraetal2008}\acb{, whether
at the inner edge of the dead zone} \citep{lyramaclow2012}, \acb{or \mordecai{at} the} outer one \wlad{\citep{lyraetal2015}}, despite the transition in 
resistivity in the outer disk being smooth \citep{dzyurkevich13}.
In regions without the steep vortensity gradient required for \acb{the} RWI, 
a vortex could be maintained by baroclinic feedback \acb{as long as (1) a} non-zero 
entropy gradient exists \citep{Klahr-Bodenheimer03,Klahr04}, \acb{(2)} the 
thermal time is finite \citep{Petersen07a,Petersen07b,Lesur-Papaloizou10,Raettig13}, 
and \acb{(3)} magnetization is absent \citep{Lyra-Klahr11}\acb{. Such conditions} may 
or may not occur in the outer regions of protoplanetary disks. 

\mordecai{T}here are also difficulties in interpreting \mordecai{gaps} as \mordecai{unambiguous} signposts of planet-disk
interactions. The transition disk surrounding HD 100546, for instance,
contains a gap, but the gap edge is \mordecai{far} too smooth to have been caused by a
planet \citep{muldersetal13}. In optically thin disks, \citet{lyrakuchner2013} 
find that \wlad{the combination of photoelectric heating and dust trapping} 
can lead to the production of sharp rings \mordecai{similar to those}
commonly attributed to gravitational  
shepherding by planets. 
\acb{Furthermore, the magnetorotational instability (MRI; Balbus \&
  Hawley 1991) could lead to gaps in the dust distribution by inducing
  zonal flows that are triggered by magnetic flux accumulation in
  turbulence \citep{lyraetal2008,johansenetal2009,simonetal2012,KunzLesur2013,flocketal2015}.}

\wlad{\acb{\mordecai{Finally, a}  one-to-one correlation between planets and spirals is also absent. 
Shear causes any density wave in a differentially rotating disk to propagate as a 
spiral.} \mordecai{Therefore}, the observed spiral features \acb{could easily} be the result of 
disk turbulence. \acb{For example, \citet{lyraetal2015}} show that waves from the MRI-active zone 
propagate into the MRI-dead zone as a coherent spiral. In fact, it has been difficult
to explain the spiral features observed in transition disk as unequivocally \acb{due to planets}.
The observed spirals have\acb{,} in general\acb{,} very large pitch \acb{angles, which is} 
inconsistent with the background disk temperature if \acb{linear spiral density wave theory for an isothermal gas}
is assumed \citep{rafikov2002,mutoetal2012}. 
\mordecai{\citet{currieetal2014} 
present a remarkable spiral feature in the disk around HD 100546, with little 
polarization, \mordecai{that can only be fit with an effective disk}
temperature of $\approx$1000\,K, substantially higher than the
expected gas temperatures. \citet{benistyetal2015} show 
that fitting a model to the spiral in the disk around MWC\,758 requires \acb{a} large disk aspect ratio 
around the spiral features, corresponding to 300\,K at 55AU, whereas the background 
gas is at much colder temperatures, 50\,K. \citet{juhaszetal2014} show 
that a local increase in pressure scale height by at least \mordecai{20\%} would be required to 
reproduce observations of multiple disks in polarized scattered
light.
}} 

\wlad{The \mordecai{high temperatures implied by these} spiral features motivate us to abandon the idea of local 
isothermality, and \mordecai{instead} \acb{to entertain potential effects
  due to the departure \mordecai{from} barotropic conditions}.  
Although numerical simulations of planet-disk interactions have 
been performed for decades, a search of the literature reveals that most 
previous works have either made use of the locally isothermal approximation 
or examined planetary masses of no more than $\sim$1--5$\Mjup$. As some of the 
candidate planets in transition disks are very massive ($\geq
10\Mjup$), we \mordecai{also examine} that region of the parameter
space. We find that the wake generated  
by \mordecai{the most} massive planets \mordecai{has relative} Mach numbers
above unity \mordecai{compared to the quiescent disk}\mordecai{. T}he resulting shocks 
heat the surrounding gas, effectively converting the planet's 
gravitational potential energy into a powerful extra energy source
\mordecai{for disk heating}. \mordecai{For cases with sufficiently long
  cooling times} \acb{(\mordecai{and in which} the thin-disk approximation \mordecai{holds}), a 
radial buoyant instability \mordecai{occurs}, leading to turbulence around the planet and} disruption 
of the usual wake features, i.e., spirals, gaps, and vortices.}

\acb{This paper is the first in a series in which we compare the
  behavior of planets embedded in
disks that are evolved with and without the assumption of local isothermality.
\mordecai{The next paper in the series} will extend the
\mordecai{two-dimensional (2D)} study
\mordecai{presented here} to include three-dimensional disks.
In Sect.~\ref{sect:model}, we present our numerical model
model\mordecai{. We} discuss the simulation results in
Sect.~\ref{sect:results}, and \mordecai{conclude} in Sect.~\ref{sect:conclusions}.}

\section{The model}
\label{sect:model}
\subsection{Governing equations}
In this work, we perform two-dimensional global hydrodynamical simulations of
gas disks with an embedded massive planet to explore the \acb{effects of shocks that result from planet-disk interactions in non-barotropic disks}. 
We \acb{primarily use the} {\sc Pencil Code}\footnote{See \url{http://pencil-code.googlecode.com}}, 
a high-order finite-difference grid \acb{hydrodynamics} code. \acb{An
  independent code, {\sc BoxzyHydro}, is also used to validate the
  main results, \mordecai{as described in \sect{sect:BoxzyHydro}}.}
\wlad{For the Pencil simulations}, the gas density is evolved using the
continuity equation

\begin{equation}
  \aderiv{\varSigma} = -\varSigma \Div \v{u} 
   \label{eq:continuity}
\end{equation}

\noindent and the equation of motion

\begin{equation}
  \aderiv{\v{u}} = -\frac{1}{\varSigma}\grad P - \grad \varPhi +
  \frac{1}{\varSigma}\Div{\v{\zeta}}, 
   \label{eq:navier-stokes}
\end{equation}

\noindent where $\varSigma$, $\v{u}$, and $P$ are the surface density, velocity,
and pressure of the gas, respectively, and $\varPhi$ represents the
gravitational potential contributions of the star and planet (we have ignored
the effects of gas self-gravity). The operator

\begin{equation}
  \aderiv{} \equiv \pderiv{ }{t} + \v{u}\cdot\del 
\end{equation} 

\noindent represents the advective derivative. The last term in the equation of
motion is the acceleration due to shock viscosity. Assuming that the shock 
viscosity \mordecai{$\nu_{\rm sh}$} takes the form of a bulk viscosity, the stress tensor is 

\begin{equation}
  \v{\zeta}_{ij}= \nu_{\rm sh} \, \varSigma \, \delta_{ij} \ \Div\v{u},
\end{equation}

\noindent 
\mordecai{The shock viscosity must} be a
localized function, ensuring that \mordecai{sufficient} energy is dissipated in regions of the flow
where shocks occur \mordecai{to satisfy the Rankine-Hugoniot jump
  conditions, while} more quiescent regions are left unaffected. 
\acb{We take}

\begin{equation}
\label{shock-visc}
  \nu_{\rm sh } = c_{\rm sh} \left<\max_3[(-\Div\v{u})_+]\right>{\left[\min(\Delta r,r\Delta \phi)\right]}^2\mordecai{.}
\end{equation}

\noindent \acb{\mordecai{T}he viscosity is} proportional to the \wlad{maximum} of positive flow convergence, \acb{ as evaluated
over \wlad{three grid cells} in each direction for a total of 9 zones in 2D (the given cell plus its immediate neighbors).} \wlad{
The angled brackets 
represent a \acb{quadratic smoothing function that smooths the divergence over seven zones in each direction, 
with weights $(1,6,15,20,15,6,1)/64$}. The result is then scaled by the square of the smallest grid spacing. 
In the above equation\acb{,} $r$ is the cylindrical radius and $\phi$ the azimuthal angle; 
$\Delta r$ and $\Delta \phi$ \mordecai{give} the numerical grid spacing.} The quantity $c_{\rm sh}$ is a constant defining the strength of the
shock viscosity, set to unity in our simulations. \wlad{Varying this coefficient 
changes the area over which the added entropy is distributed but not the total amount 
of extra energy, \mordecai{which still must satisfy the jump
  conditions.}} We refer to \acb{$c_{\rm sh}$} as the shock viscosity 
coefficient \citep[for more details see][]{haugenetal2004}. 

To follow the thermodynamic evolution of the disk, we assume the ideal gas law
as the equation of state

\begin{equation}
  P = \varSigma \cs^2/\gamma,
\end{equation}

\noindent where $\cs$ is the \acb{adiabatic} sound speed and $\gamma$ is the adiabatic index. 
The temperature and sound speed are related according to 

\begin{equation}
  T=\cs^2/[c_p(\gamma-1)],
\end{equation} 

\noindent where $c_p=\gamma\cv$ is the specific heat at constant pressure\wlad{, and 
$\cv$ the specific heat at constant volume. They are related to the universal gas constant $R$
by \[R/\mu=c_p-\cv,\] where $\mu$ is the mean molecular weight of the gas}.

The Pencil Code implements thermodynamic evolution by solving for the entropy
$S=\cv\left(\ln P - \gamma\ln\varSigma\right)$. We use thermal relaxation for non-isothermal runs, driving
the temperature to a reference temperature $T_{\rm ref}$ within a time $\taut$,
according to 

\begin{equation}
   T  \aderiv{S} = - \cv \frac{\left(T-T_{\rm ref}\right)}{\taut} + \varGamma_{\rm sh},
   \label{eq:entropy-equation}
\end{equation}

\noindent where  both $T_{\rm ref}$ and $\taut$ are radially-varying.  We take
$T_{\rm ref}$ to be the initial temperature and  $\taut$ proportional to the
orbital period

\begin{equation}
  \taut = \tauref \ \varOmega_0/\varOmega,
\end{equation}

\noindent where  $\varOmega=\sqrt{GM_\star/r^3}$ is the Keplerian angular
frequency.  The subscript ``0'' refers to the values of the quantities at an
arbitrary reference radius $r_0=1$. The reference timescale for thermal
relaxation, $\tauref$, is a free parameter of the model reflecting the
radiative cooling timescale of the gas. The function 

\begin{equation}\label{eq:shocktemp}
  \varGamma_{\rm sh} = \nu_{\rm sh} \left(\Div\v{u}\right)^2
\end{equation} is the heating due to dissipation of shocks. 

Sixth-order hyper-dissipation terms are added to the evolution equations to
provide extra dissipation near the grid scale, \acb{as discussed in}
\citet{lyraetal2008}. \acb{These terms} are needed \mordecai{for
  numerical stability} because the high-order scheme of the
Pencil Code 
\mordecai{formally lacks error terms to produce stabilizing}
 numerical dissipation \citep{mcnallyetal2012}.

The simulations are run in the inertial frame, centered at the barycenter of the 
system. The positions of the star and planet are evolved on circular
orbits \mordecai{ by direct integration} using an
$N$-body routine, employing the same 3rd-order Runge-Kutta algorithm 
used for updating the gas grid. Since we are modeling disks with
high-mass protoplanets, this will allow us to capture any dynamical effects on
the gas induced by stellar wobble. The gravitational potential $\varPhi$ acting
on the star, planet, and gas is the sum of the potential contributions of the
star and planet (i.e. the gravity of the gas is ignored):
\begin{equation}
  \varPhi = -\sum_i^N \frac{GM_i}{\sqrt{\mathcal{R}_i^2 + b_i^2}},
  \label{eq:potential}
\end{equation}
where $N=2$, $M_i$ is the mass of particle $i$,
$\mathcal{R}_i=\left|r-r_i\right|$ is the distance between the $i$th particle
and a gas parcel or massive particle for which the potential is being
calculated, and $b_i$ is the smoothing radius of the $i$th particle\acb{, which is used to prevent singularities.}
 The gravitational smoothing radius of the planet
is set to \mordecai{its} Hill radius\acb{. T}he star does not require a smoothing radius
\acb{because} it lies outside the gas grid.

For units, the planetary semi-major axis and orbital velocity are used as the
units of length and velocity, respectively. This in turn implies 2$\pi$ time
units per orbit.

\subsection{Run parameters}
All simulations are performed on a cylindrical grid with ($N_r$,~$N_\theta$) =
(768,~1024), with 0.4 < $r$ < 12 \mordecai{(in units of \acb{the planet's} semi-major axis)}. Grid points are spaced in $r$ according to a
power law such that $\Delta r \propto r^{0.5}$, providing higher resolution in
the vicinity of the planet compared with the outer disk 
\wlad{(for our choice of initial temperature gradient, this non-equidistant grid corresponds to 
a constant number of radial grid cells per scale height)}. For all simulations,
the initial density distribution is axisymmetric and decreases with the square
root of the radius, with $\varSigma_0=1$ (the physical units of surface density
can be chosen arbitrarily since the self-gravity of the gas has not been
included). The sound speed at the reference radius was set to $\xtimes{5}{-2}$,
which corresponds to a temperature of \textasciitilde100\,K, assuming a
1\,$\Msun$ star with planetary semi-major axis 5.2\,AU, $\gamma=7/5$,
and mean molecular weight 2.4 (corresponding to a 5:2 hydrogen/helium
mixture by mass).
The initial sound speed is set to
\begin{equation}
  \cs^2 = \csO^2 r^{\alphat} . \label{eq:tempslope}
\end{equation}

In order to study the role of thermodynamical evolution in disks containing
high-mass protoplanets, we perform four sets of simulations. First, we perform
simulations for three planet-to-star mass ratios $q=\ttimes{-3}$,
$\xtimes{5}{-3}$, and $\ttimes{-2}$, with locally isothermal and adiabatic disks.
We run one adiabatic simulation \wlad{\acb{with shock heating artificially turned off} 
in the entropy equation} ($\varGamma_{\rm sh}=0$) \wlad{\acb{,} while keeping the acceleration due 
to shock viscosity \acb{turned on} in the momentum equation\acb{.}} \acb{This is done to determine the role of shock heating} 
in the non-barotropic simulations. \acb{For this test case,} $q=\ttimes{-2}$. Next, we run three simulations with
different values of the reference thermal relaxation timescale $\tauref$\acb{, also} with
$q=\ttimes{-2}$.  Finally, we run three \acb{additional simulations} with mass ratio
$\ttimes{-2}$\acb{, but} using different radial power laws for the initial temperature
(all adiabatic).  Table \ref{tab:runs} contains run parameters for all
simulations, thirteen in total, where $q$ is the planet-to-star mass ratio,
$\alphat$ is the slope of the radial temperature power law, $\gamma$ is the
adiabatic index of the gas, and $\tauref$ is the cooling timescale in units of
orbits at the reference radius.  Locally isothermal runs are denoted by
$\tauref=0$ and $\gamma=1$, while adiabatic runs are denoted by
$\tauref=\infty$. All simulations are run for 100 planetary orbits.

\begin{table}
\caption{Simulation parameters}
\begin{center}
\begin{tabular}{lccccc}
\hline
\multirow{2}{1em}{Run}     & \multirow{2}{1em}{$q$}  & \multirow{2}{1em}{$\alphat$}  & \multirow{2}{1em}{$\gamma$}   & \multirow{2}{1em}{$\tauref$}  &  Shock   \\
                           &                         &                               &                               &                                 & heating \\
\hline
\multicolumn{6}{c}{Mass, thermal evolution; \Figss{fig:massentropy}{fig:shockcheck}} \\
A                          & $\ttimes{-3}$           & -1                            & 1                             & 0                               &  ---     \\
B                          & $\xtimes{5}{-3}$        & -1                            & 1                             & 0                               &  ---     \\
C                          & $\ttimes{-2}$           & -1                            & 1                             & 0                               &  ---     \\
D                          & $\ttimes{-3}$           & -1                            & 1.4                           & $\infty$                        &  Yes     \\
E                          & $\xtimes{5}{-3}$        & -1                            & 1.4                           & $\infty$                        &  Yes     \\
F                          & $\ttimes{-2}$           & -1                            & 1.4                           & $\infty$                        &  Yes     \\
\hline
\multicolumn{6}{c}{Viscous heating; \fig{fig:shockheatingcomp}} \\
G                          & $\ttimes{-2}$           & -1                            & 1.4                           & $\infty$                        &  No      \\
\hline
\multicolumn{6}{c}{Relaxation time; \fig{fig:coolingtime}} \\
H                          & $\ttimes{-2}$           & -1                            & 1.4                           & 0.1                             &  Yes     \\
I                          & $\ttimes{-2}$           & -1                            & 1.4                           & 1                               &  Yes     \\
J                          & $\ttimes{-2}$           & -1                            & 1.4                           & 10                              &  Yes     \\
\hline
\multicolumn{6}{c}{Temperature power law; \fig{fig:tempslope}} \\
K                          & $\ttimes{-2}$           & -0.5                          & 1.4                           & $\infty$                        &  Yes     \\
L                          & $\ttimes{-2}$           & -0.2                          & 1.4                           & $\infty$                        &  Yes     \\
M                          & $\ttimes{-2}$           &  0                            & 1.4                           & $\infty$                        &  Yes     \\
\hline
\end{tabular}
\end{center}
\label{tab:runs}
\end{table}

\section{Results}
\label{sect:results}
\subsection{Global disk properties}
\subsubsection{Planet mass and thermal evolution}
\begin{figure*}
  \begin{center}
    \resizebox{\textwidth}{!}{\includegraphics{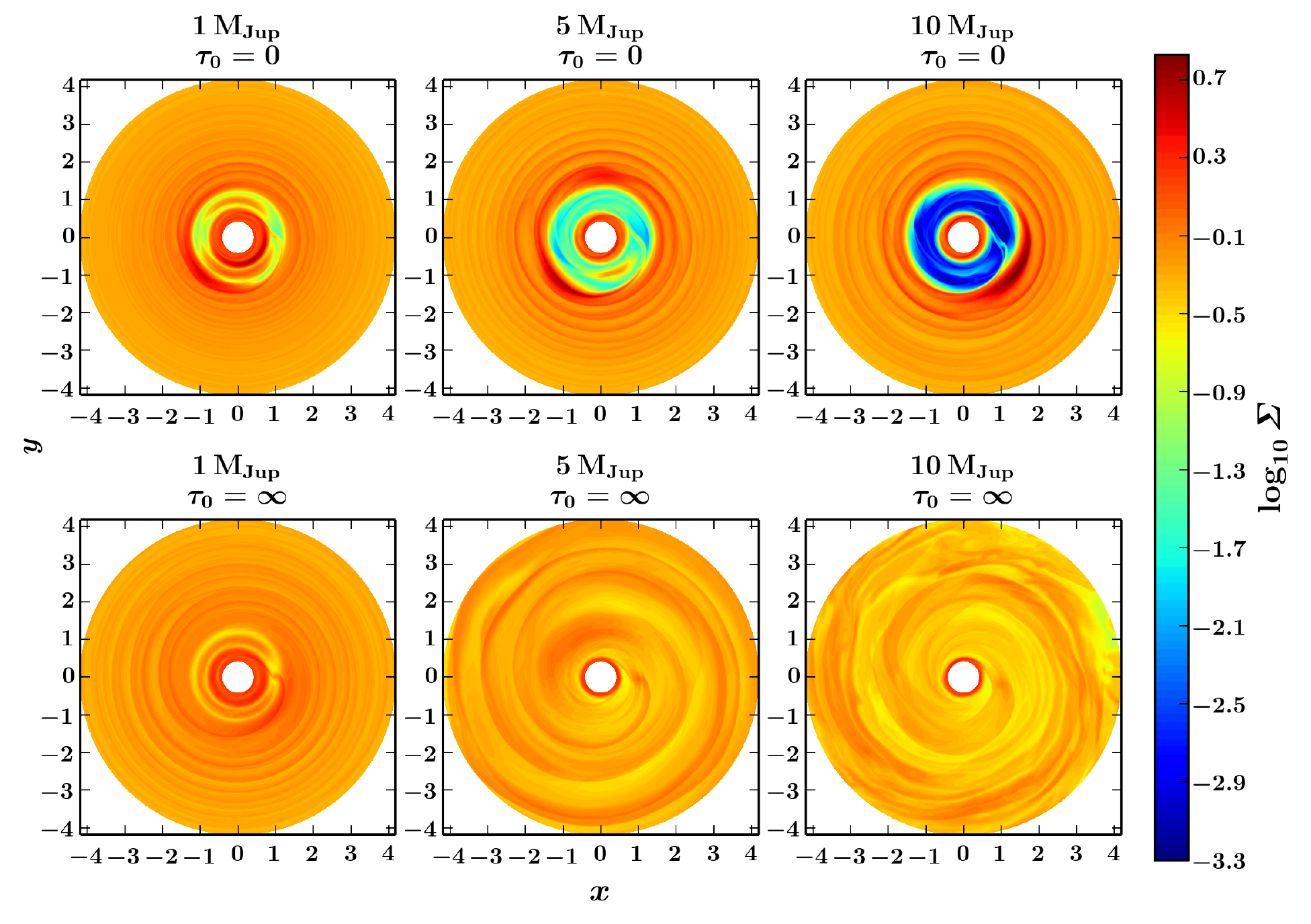}}
    \vspace{\figspace}
  \end{center}
  \caption[]{Surface densities in the inner disk ($r<4$) after 100 orbits for
    locally isothermal ($\tauref=0$) runs A, B, and C, and adiabatic
    ($\tauref=\infty$) runs D, E, and F. Turbulence due to buoyancy appears in
    adiabatic runs and increases with $q$ (masses based on
    M\,$_\mathrm{star}\approx1\,\Msun$).}
  \label{fig:massentropy}
\end{figure*}

\Fig{fig:massentropy} shows  \acb{the} surface densities in the inner disk  \acb{($r<4$)} after 100 orbits
for \acb{the} locally isothermal ($\tauref=0$) and adiabatic ($\tauref=\infty$) runs
\acb{and} for three values of \acb{the} planet-star mass ratio $q$.
\acb{For all $q$, there is a clear difference between the global morphologies of the isothermal and adiabatic disks.
All three isothermal simulations show well-behaved spiral structure, a clear gap, and an accompanying vortex, as expected.
In contrast, large-scale turbulence is clearly present in the $q=\xtimes{5}{-3}$ and $\ttimes{-2}$ adiabatic runs, in which the expected gaps and vortices appear to be completely missing.
For the $q=\ttimes{-3}$ adiabatic run, the presence of global turbulence is not as apparent,  but the gap depth and the accompanying vortex are visibly diminished.}

Since the turbulence seen in \Fig{fig:massentropy} is present only in runs with
non-barotropic equation of state, we hypothesize that it is the result of
buoyant oscillations triggered by hot gas that has been tidally compressed 
by the planet. 

To test this hypothesis and confirm the
originating region of any such oscillations, we compare the results of each
simulation to the Solberg-H{\o}iland stability criterion. The
criterion for hydrodynamic stability is given by 
\begin{equation}
  \kappa^2 + N^2 \geq 0,
  \label{eq:sh}
\end{equation}
where 
the epicyclic frequency
\begin{equation}
  \kappa \equiv \left[\frac{1}{r^3} \deriv{}{r} \left(r^4  \dot\theta^2\right)\right]^{1/2},
\end{equation}  and 
the characteristic frequency of buoyant oscillations,
the Brunt-V{\"a}is{\"a}l{\"a} frequency, 
\begin{equation}
  N \equiv \left[\frac{1}{\varSigma} \deriv{P}{r} \left( \frac{1}{\varSigma} \deriv{\varSigma}{r} - \frac{1}{\gamma P} \deriv{P}{r} \right)\right]^{1/2}.
\end{equation}
In the case of a steep entropy gradient, $N^2$ will become negative, leading to
buoyant instability. For positive $\kappa^2$, the velocity shear of the disk
will tend to damp buoyant oscillations.  We calculate $\varLambda$ for each
simulation at several timesteps, where
\begin{equation}\label{eq:lambda}
  \varLambda \equiv 1 + \frac{N^2}{\kappa^2} 
\end{equation}
is the left-hand side of Eq\mordecai{uation\,(}\ref{eq:sh}), normalized by $\kappa^2$. For
$\varLambda < 0$, the local shear of the disk will not stabilize buoyant
oscillations. If the turbulence observed in \Fig{fig:massentropy} is due to
buoyant instabilities, then $\varLambda$ will achieve negative values \mordecai{only} in the
adiabatic simulations. This is because 
   in the locally isothermal simulations neither the temperature profile
(\mordecai{which remains} constant through the simulation) nor
\mordecai{the} density fluctuations are sufficient to
give rise to a steep entropy gradient.

\begin{figure}
  \begin{center}
    \resizebox{.495\columnwidth}{!}{\includegraphics{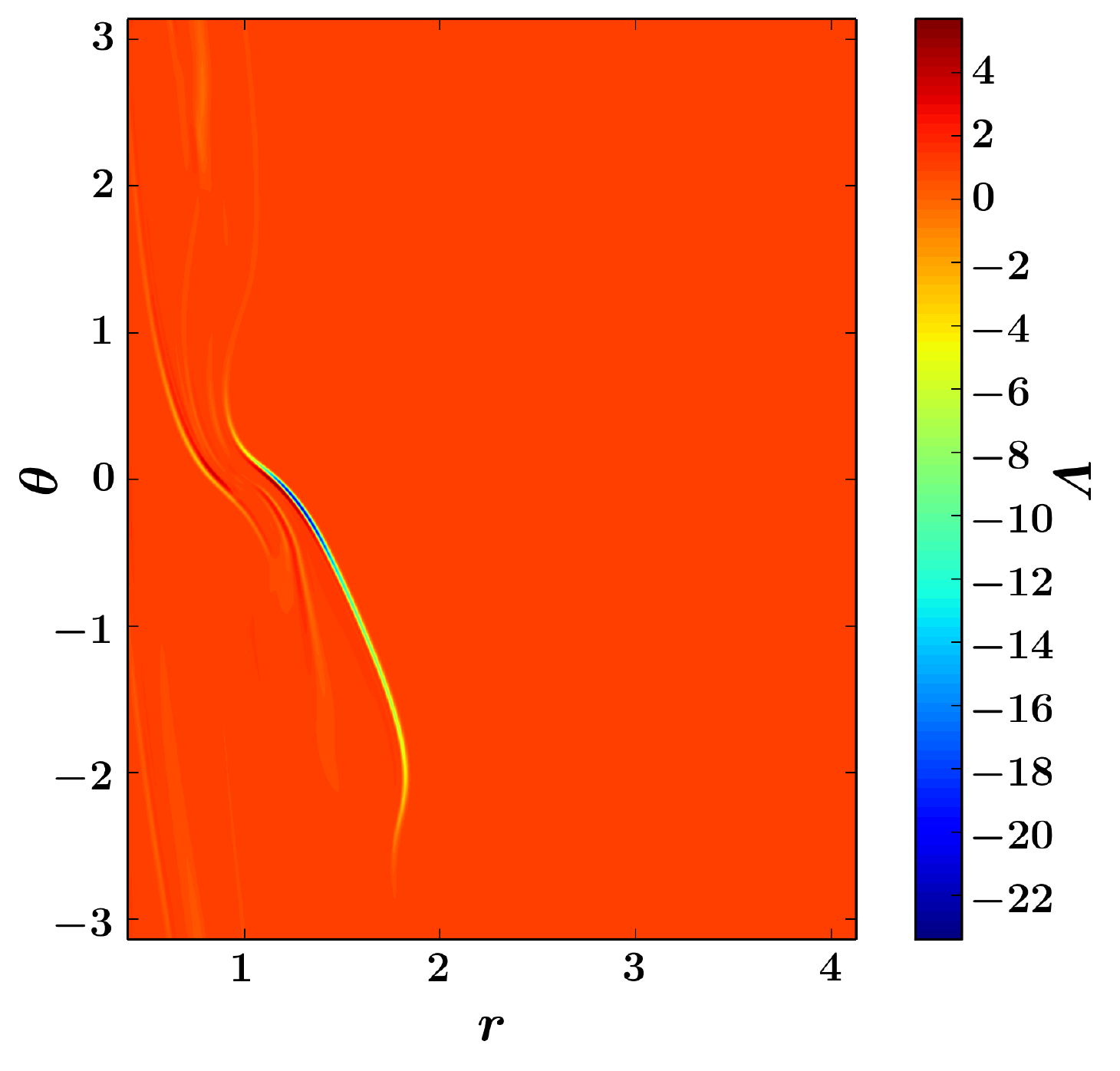}}
    \resizebox{.495\columnwidth}{!}{\includegraphics{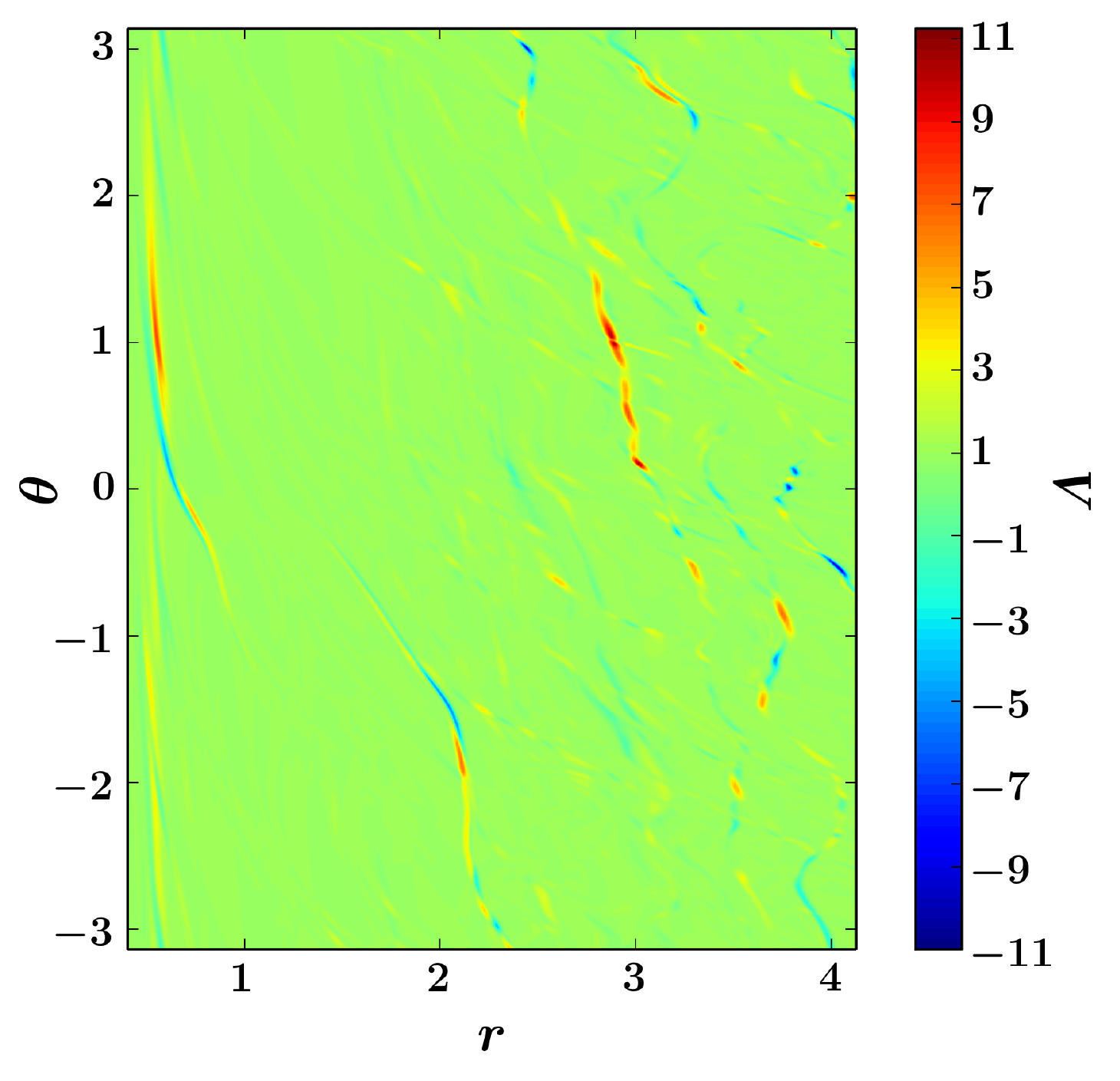}}
    \vspace{\figspace}
  \end{center}
  \caption[]{$\varLambda$ for $q=\ttimes{-2}$ after 1 orbit (left 
    panel) and 100 orbits (right panel) for the \mordecai{inner disk ($r
      < 4$) in the} adiabatic case
    (\mordecai{run }F). Buoyant instability \mordecai{occurs for $\varLambda < 0$.  It}
    begins in the outer spiral wake 
    produced by the planet after 1 orbit, \mordecai{while} instability is triggered
    in the inner spiral wake and outer disk soon after.

}
  \label{fig:lambda10mjup}
\end{figure}

For the locally isothermal simulations, we indeed find that
$\varLambda$ never 
\mordecai{becomes negative}. In the adiabatic case, on the other hand, for $q=\ttimes{-2}$
(run F), $\varLambda$, shown in left panel of \Fig{fig:lambda10mjup}, reaches
negative values after less than one orbit in the region of the outer spiral wake
of the planet, as well as, to a lesser extent, the inner spiral wake. After 100
orbits, zones of buoyant instability are seen throughout the disk, as shown in the
right panel of \Fig{fig:lambda10mjup}. For the $q=\ttimes{-3}$ case (run D),
$\varLambda$ achieves negative values for the first \textasciitilde15 orbits,
then stays positive for the remainder of the simulation, explaining the relative
similarity in resulting surface densities between the \acb{corresponding}  isothermal and adiabatic
cases \acb{(runs A and D, respectively)}. These results confirm that buoyant instabilities are \acb{consistent with} the large-scale \wlad{disruption of the usual planet-disk interaction features} seen in the lower
panels of \Fig{fig:massentropy}.

\subsubsection{Shock heating}
\begin{figure*}
  \begin{center}
    \resizebox{\textwidth}{!}{\includegraphics{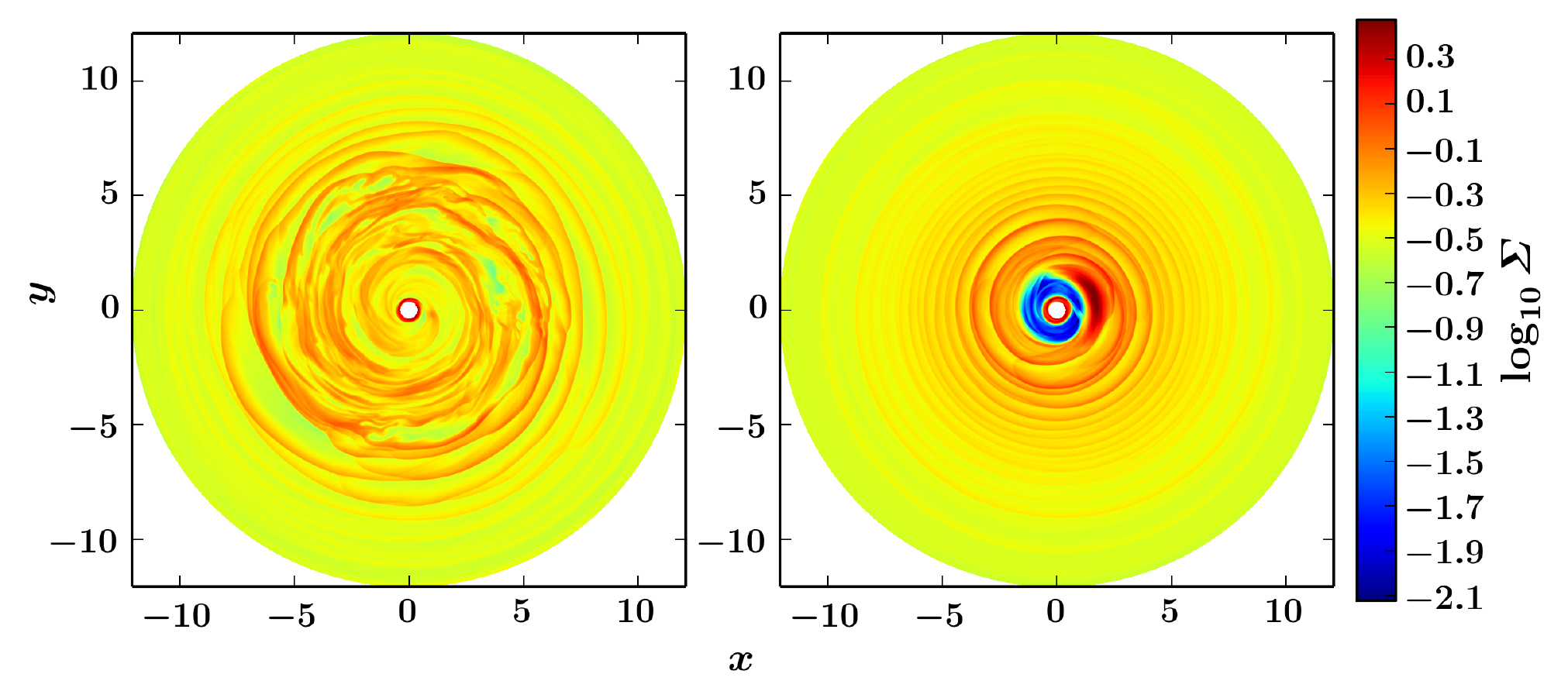}}
    \vspace{\figspace}
  \end{center}
  \caption[]{Adiabatic simulation of planet-disk interaction of a 10$\Mjup$ planet (left) with shock heating and (right) without shock heating. This test conclusively shows that the energy dissipated in shocks is driving the instability.} 
  \label{fig:shockheatingcomp}
\end{figure*}

In order to determine the role of shock heating in generating the observed
buoyant instability, we run another simulation with an adiabatic disk and
a 10$\Mjup$ planet where we exclude heating of the gas due to shocks
($\varGamma_{\rm sh} = 0$; run G).  The resulting surface density after 100 orbits,
shown in the right panel of \Fig{fig:shockheatingcomp}, resembles that of the locally isothermal
case, where the gap and vortex are visible. \wlad{This test conclusively shows that 
the entropy generated in shocks is responsible for the novel phenomen\mordecai{a} described 
in the present work.}

\subsubsection{\wlad{How strong are the shocks?}}

\begin{figure}
  \begin{center}
    \resizebox{.495\columnwidth}{!}{\includegraphics{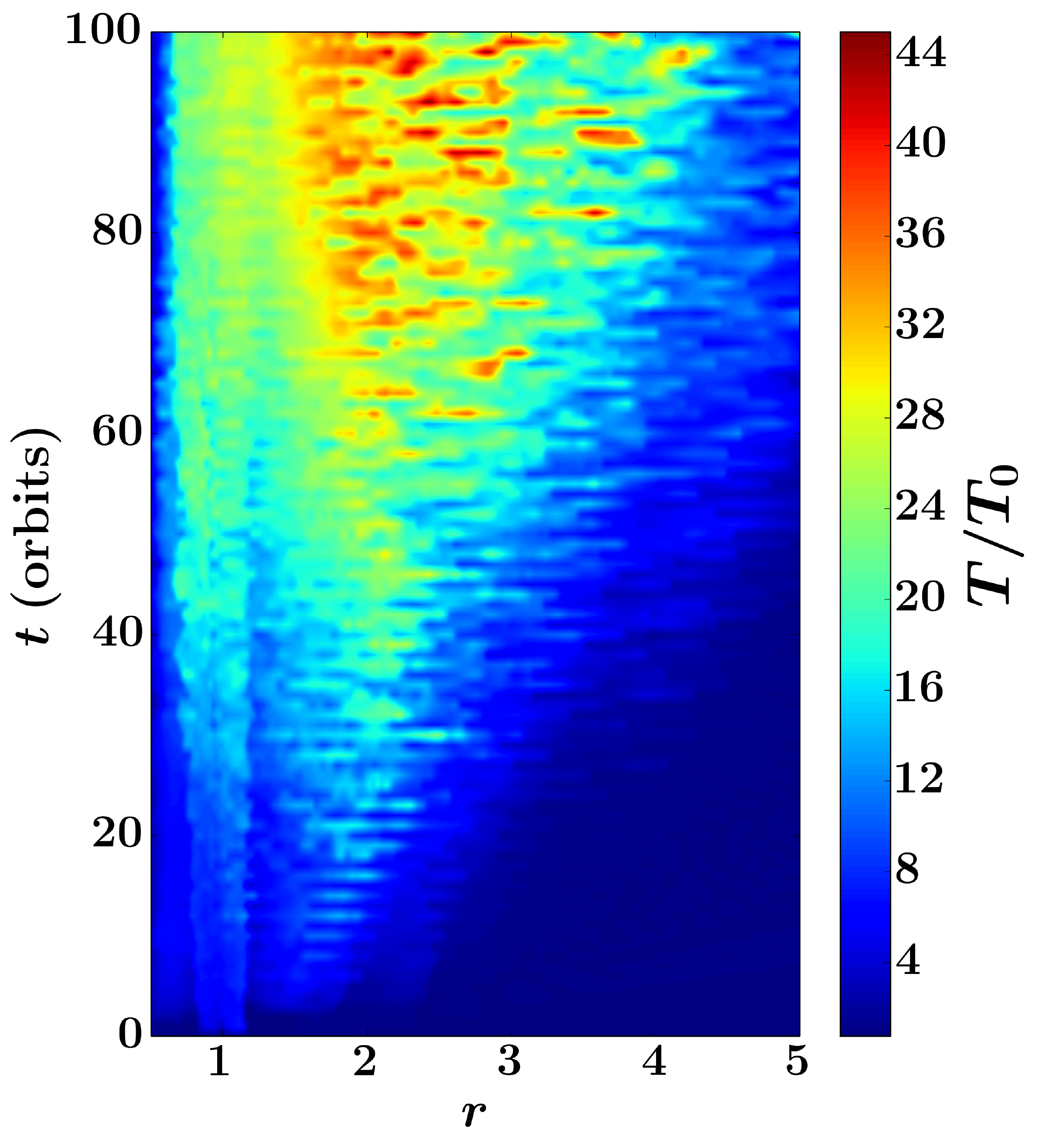}}
    \hspace{-1mm}
    \resizebox{.495\columnwidth}{!}{\includegraphics{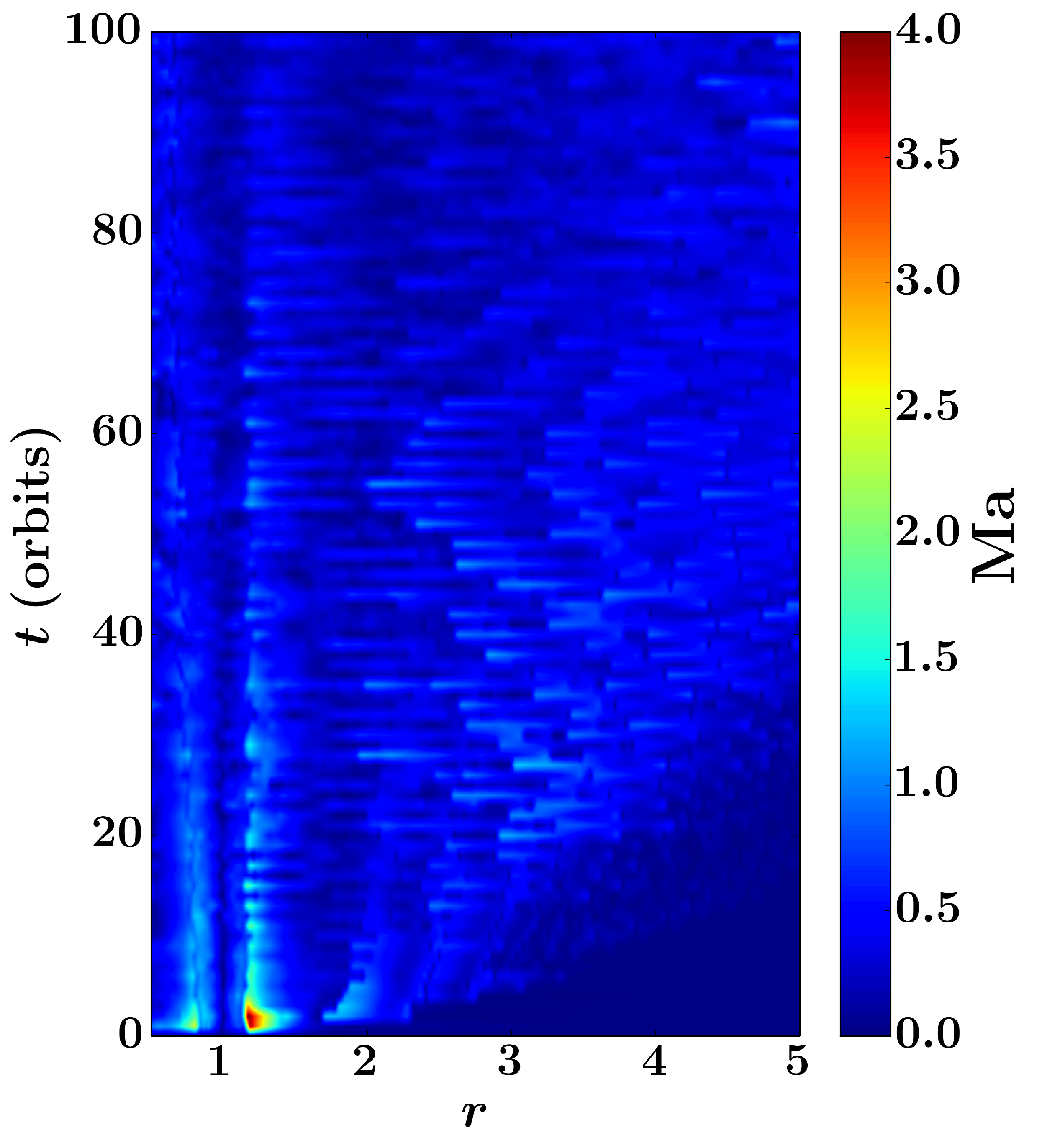}}
    \vspace{\figspace}
  \end{center}
  \caption[]{Temperature and Mach number for \mordecai{the inner disk ($r
      < 5$) of} run F, at the azimuthal position of
the planet. Heating due to shocks initially occurs close to the planet, but
spreads to the rest of the disk over the course of the simulation.
}
  \label{fig:shockcheck}
\end{figure}

In \Fig{fig:shockcheck} we show radial profiles of temperature (normalized
by the initial temperature profile) and Mach number\acb{ $\Ma$} for 100 orbits in the
adiabatic, 10$\Mjup$ case \acb{(run F)} at the azimuthal position of the planet. Shocks
propagate throughout the disk for the duration of the simulation\acb{;} however\acb{,} they
do not grow rapidly and monotonically like the temperature, \acb{with} the strongest
shocks ($\Ma>1.5$) \acb{subsiding} after $\sim50$ orbits. The temperature
increases steadily throughout the disk over the course of the simulation. \wlad{This 
shows that the temperature increase is not due to a single shock, but due to the 
entropy injection of \mordecai{multiple} shocks, over the course of many orbits.} 

\subsubsection{Thermal relaxation timescale}

\begin{figure}
  \begin{center}
    \resizebox{\columnwidth}{!}{\includegraphics{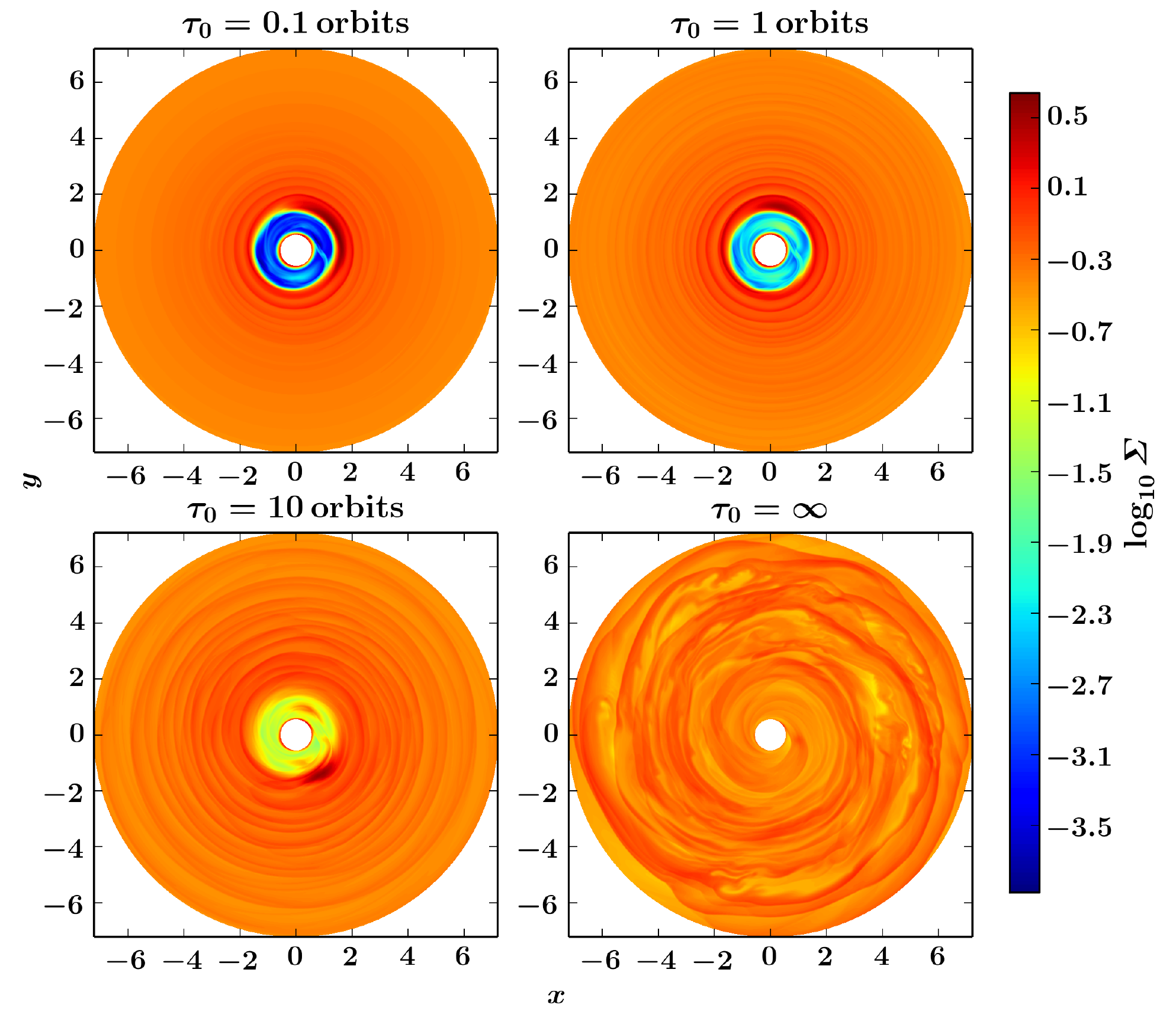}}
    \vspace{\figspace}
  \end{center}
  \caption[]{Surface densities \mordecai{for the inner half of the disk
      ($r < 6$)} after 100 orbits for four different cooling
   timescales (runs H, I, J, and F)\mordecai{, for a planet with $q = \ttimes{-2}$}.}
  \label{fig:coolingtime}
\end{figure}

To investigate the importance of cooling in non-barotropic disks with
high-mass planets, we run several simulations with $q=\ttimes{-2}$ for
\mordecai{cooling times}
$\tauref$ = 0.1, 1, 10, and $\infty$ orbits (runs H, I, J, and F,
respectively). Surface densities after 100 planetary orbits are shown in
\Fig{fig:coolingtime}. Not surprisingly, simulations with shorter relaxation
times (0.1 and 1 orbits) more closely resemble the isothermal case\mordecai{:} buoyant
instabilities and subsequent turbulence are minimal, \mordecai{and so} the gap and
vortex are preserved\mordecai{. F}or longer cooling timescales (10 orbits and
adiabatic), the shock heating drives buoyant instabilities, generating
turbulence that dominates the global structure of the disk, disrupting the gap
and vortex.

Because the wake is stationary in the reference frame of the 
planet, a parcel of gas will be excited by one of the spirals 
once per synodic period $\tau_{\rm syn}$ (with respect to the planet). Therefore, 
for $\tauref<\tau_{\rm syn}$, the parcel will cool radiatively before 
encountering the next shock, and thus the energy
contributions of successive shocks will not continue to increase the
temperature of the parcel beyond that of a single shock. For
$\tauref>\tau_{\rm syn}$, however, the gas parcel will not have time to cool to its
original temperature before encountering the next shock, therefore the
temperature of the gas will continue to rise, even though the 
    energy
contributions of each shock will be the same as in the low $\tauref$ case
(since \acb{Eq.~\ref{eq:shocktemp}} does not depend on cooling timescale). The emergence of
the observed buoyant instabilities therefore depends on the relative
orbital and cooling timescale. Its strength depends additionally on the
    energy
contribution of each shock.  These are a function of the characteristic
   differential velocity
of the gas flows associated with the spiral wakes, which is in turn
determined by the planet's mass. 

\subsubsection{Temperature power law}
Because \acb{the} \wlad{buoyant} instability results from the presence of a strong
entropy gradient, we compare simulations with differing initial temperature
power law slopes $\alphat$ to determine whether a steeper initial entropy
gradient enhances the buoyant instabilities generated
locally by the planet. We once
again use $q=\ttimes{-2}$ and an adiabatic disk, and conduct runs with $\alphat$
= 0, -0.2, -0.5, and -1, corresponding to runs M, L, K, and F, respectively.
The value for run L was chosen to flatten the initial entropy gradient, i.e.\acb{,}
$ds/dr=0$. For $\alphat=0$, $\varLambda$ is initially positive throughout the 
disk, \mordecai{while it} becomes more negative for decreasing $\alphat$,
\mordecai{so that for these values} the disk
is \mordecai{initially} slightly Solberg-H{\o}iland unstable (see Eq.~\ref{eq:lambda}).

\begin{figure}
  \begin{center}
    \resizebox{\columnwidth}{!}{\includegraphics{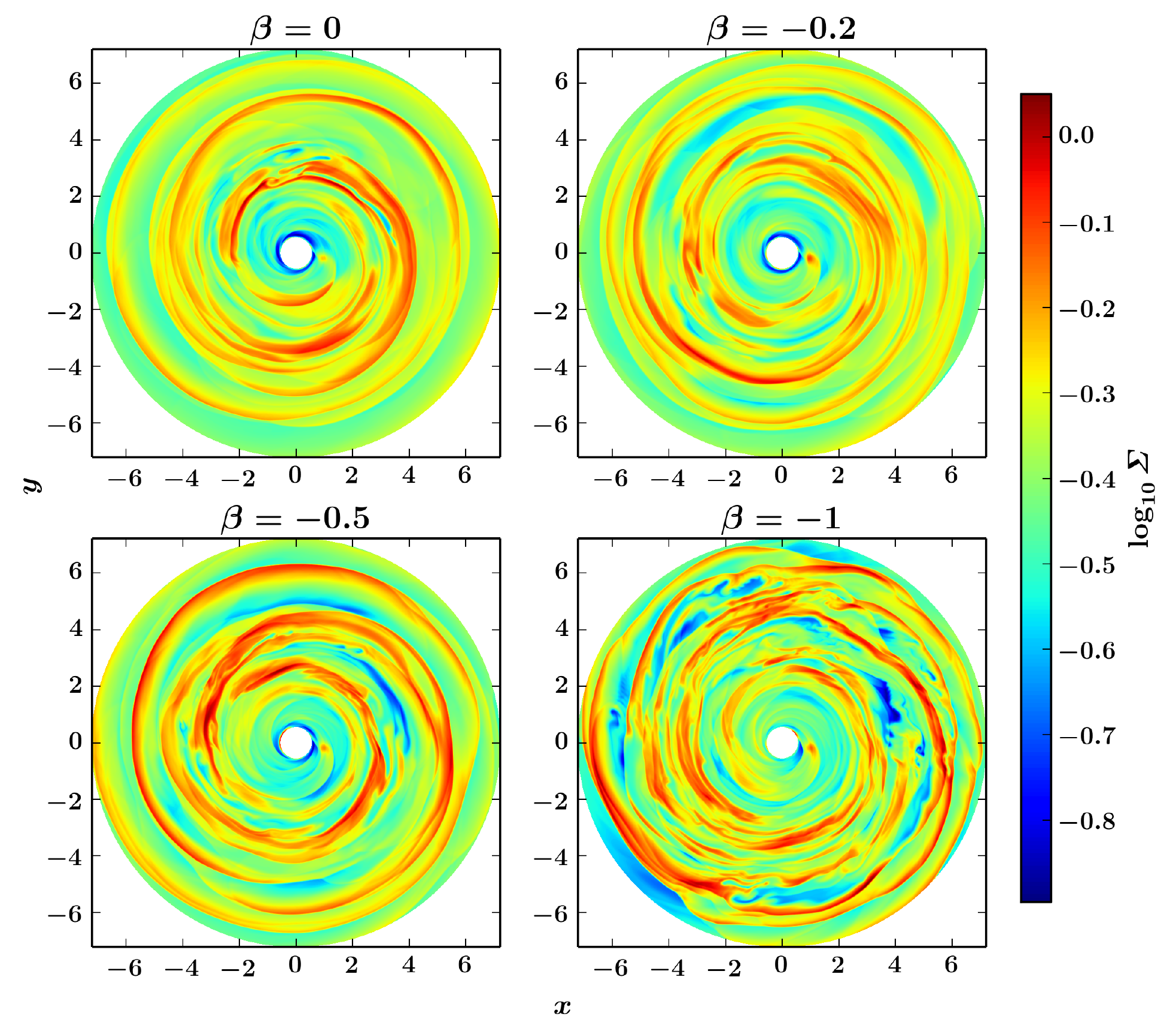}}
    \vspace{\figspace}
  \end{center}
  \caption[]{Surface densities \mordecai{for the inner half of the disk
      ($r < 6$)} after 100 orbits for four different temperature
   power law slopes (runs M, L, K, and F).
     Run M is initially globally stable according to the
     Solberg-H{\o}iland criterion, but still experiences 
     instability due to \mordecai{heating from the} planetary wakes (see Fig.~\ref{fig:lambda10mjup}).
}
  \label{fig:tempslope}
\end{figure}

The resulting surface densities after 100 planetary orbits for these simulations
are shown in \Fig{fig:tempslope}. A steep temperature slope (\mordecai{e.g.,}~$\alphat=-1$)
appears to lead to stronger turbulence, though significant turbulence is still
observed for $\alphat=0$.
The enhancement of the turbulence when the disk is globally unstable
supports our hypothesis that the planet generates turbulence by
driving local regions into instability, even when the disk is
globally stable.

\subsection{Angular momentum transport}
In order to examine angular momentum transport in the disks simulated 
in this work and thereby identify possible implications for disk evolution, 
we calculate effective \citet{shakurasunyaev1973} $\alpha$-viscosities over time for simulations of 
locally isothermal and adiabatic disks with $q=\ttimes{-2}$ (runs C and F). 
We define an effective viscosity $\nu$ in terms of $\alpha$:
\begin{equation}
  \nu=\alpha \cs^2 / \varOmega,
  \label{eq:shakura-sunyaev}
\end{equation}
which drives an accretion rate \acb{(assuming a steady state)}
%
%
\begin{eqnarray}
  \dot{m}&=&\acb{-}\wlad{2\pi\nu\varSigma \ \frac{\partial\ln\varOmega}{\partial\ln r},\nonumber}\\
        &=&3\pi\nu\varSigma\wlad{.}
\end{eqnarray}

\wlad{We define also the $\alpha$ value as a function of the kinetic stress $\delta u_r \delta u_\phi$ (where 
$\delta u_i = u_i - \langle u_i \rangle$, and $\langle X \rangle$
represents the azimuthal average of $X$) 
}

\begin{eqnarray}
  \wlad{\alpha} &\wlad{\equiv}& \wlad{\frac{2}{3}} \wlad{\frac{\langle \delta u_r ~ \delta u_\phi \rangle}{\langle c_s^2 \rangle}} \nonumber \\
  & \wlad{=} & \wlad{\frac{2}{3} \left(\frac{\langle u_r u_\phi \rangle - \langle u_r \rangle \langle u_\phi \rangle  }{\langle c_s^2 \rangle}\right)}.
\end{eqnarray}

\begin{figure}
  \begin{center}
    \resizebox{.495\columnwidth}{!}{\includegraphics{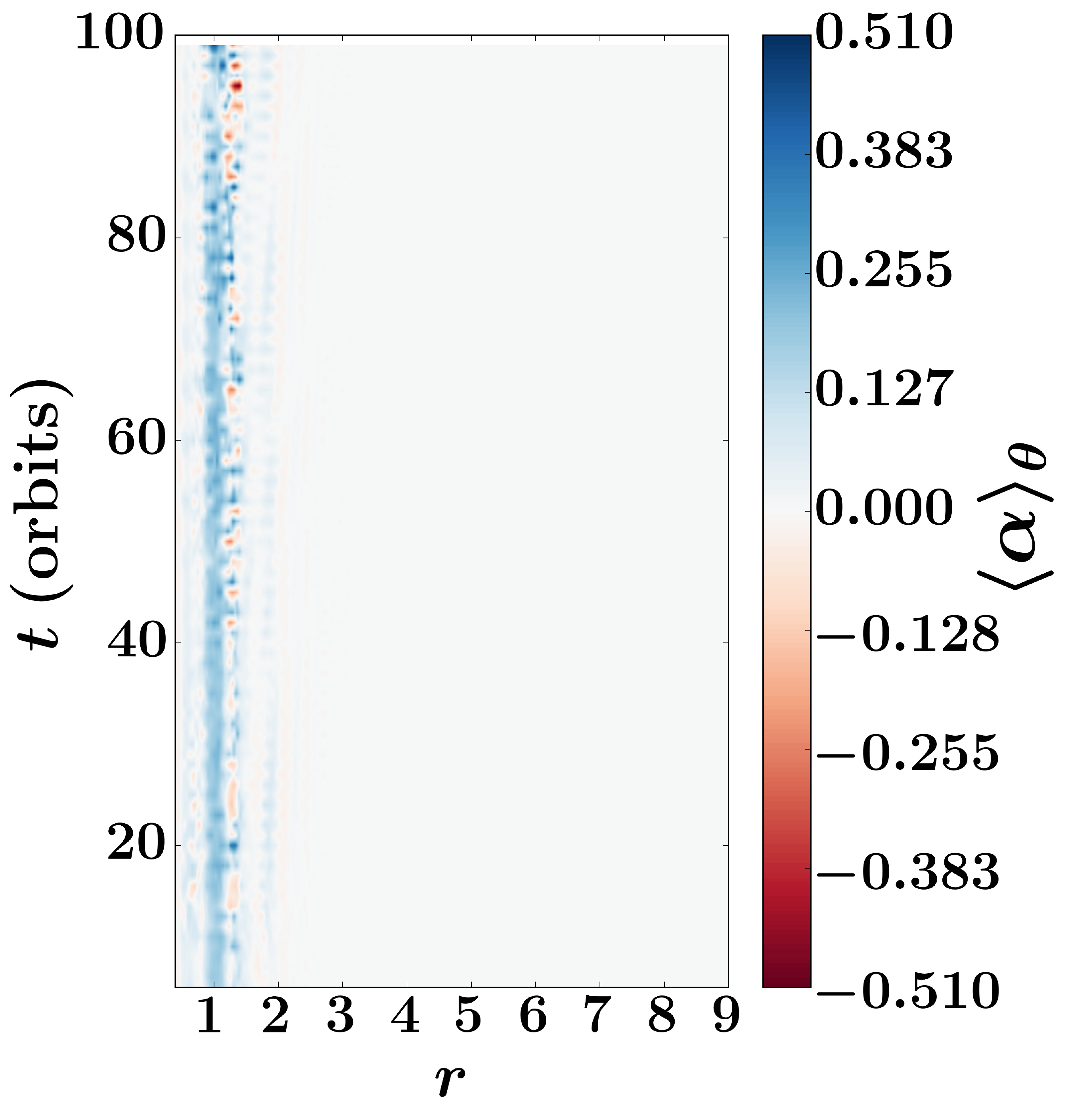}}
    \hspace{-1mm}
    \resizebox{.495\columnwidth}{!}{\includegraphics{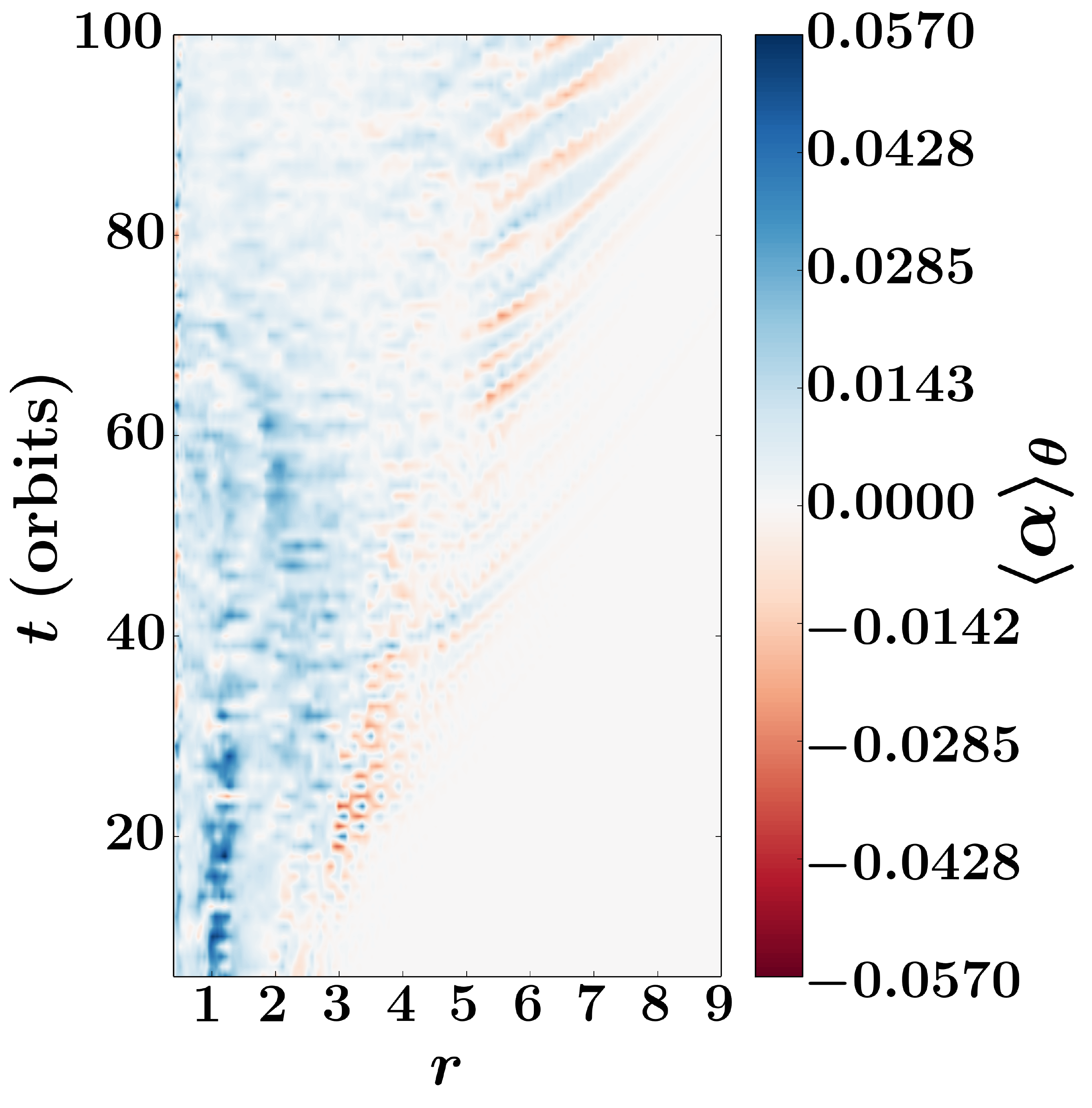}}
  \end{center}
  \caption[]{Azimuthally-averaged effective $\alpha$-viscosities for
    locally isothermal (C; left) and adiabatic (F; right)
    runs, over the course of 100 planetary orbits\mordecai{, for $r < 9$}.
    \wlad{The high $\alpha$ values in the isothermal case \acb{, taken at face value,} 
      imply nearly supersonic turbulence,
    \acb{and indicate that the local $\alpha$-disk prescription does not properly 
      describe the disk's evolution and that the thin-disk approximation may be breaking down.}
    The thin-disk approximation also limits the adiabatic case: in 
      \mordecai{three dimensions} the shocks will be weaker as vertical expansion is
      allowed. The results nevertheless \acb{highlight} that shocks in the wake\mordecai{s} 
      of high-mass planets are important for mass redistribution.}} 
  \label{fig:spacetimealphas}
\end{figure}

\Fig{fig:spacetimealphas} shows \wlad{$\alpha$} for the 
aforementioned simulations, calculated once per orbit for 100 orbits.

We find that the evolution of the disk \mordecai{driven} by \mordecai{the}
effective $\alpha$ \mordecai{viscosity differs}
substantially between the locally isothermal and adiabatic cases. 
\wlad{For the isothermal case (left panel of \Fig{fig:spacetimealphas})
the $\alpha$ values are large (reaching 0.5) but occur close to the 
planet and gap, while the disk beyond $r=$2--3 remains relatively 
quiescent.  By contrast, the turbulence in the adiabatic case
propagates outward through the disk throughout the entire simulation. The 
$\alpha$ in this case are lower, reaching 0.05, an order of magnitude less. 
The globally-averaged $\alpha$ between \alex{60--100} orbits is $\approx$0.01--0.02.} 
The relatively large value of $\alpha$ indicates rapid inward motion 
of the gas. Defining the accretion timescale
\begin{equation}
 \acb{ \tau_\mathrm{acc} \wlad{\equiv m/\dot{m}} \sim \varOmega r^2 (c_s^2 \alpha)^{-1} ,}
\end{equation}
we find that for a 10\,M$_\mathrm{Jup}$ planet orbiting a 1\,$\Msun$ star at
5.2\,AU, the characteristic accretion timescale of the disk is on the order of
\wlad{$10^5$} years.

\wlad{\acb{We} caution that this particular result pertaining \acb{to} accretion 
is a function of the cooling time, as for gravitational instability 
(\citealt{durisenetal2007,Meru&Bate2011,Meru&Bate2012})\acb{, or any shock-driven instability.} }
\wlad{The high $\alpha$ values in the isothermal case \acb{($\approx \wlad{0.5}$) 
if taken at face value, imply} that the rms velocity of the turbulence is approaching 
supersonic values ($v_{\rm rms} \sim \sqrt{\alpha}~c_s \approx 0.7 c_s$). \acb{However, 
this is mainly an indication that the local $\alpha$-disk prescription does not 
properly describe the disk's evolution and that the thin-disk approximation may be 
breaking down.} In the adiabatic case, even though the effective $\alpha$ values are reasonably 
subsonic, the large-scale turbulence is similarly implicated. In three-dimensional runs 
we expect the shocks to be weaker because of the extra degree of freedom (vertical 
expansion \acb{such as in shock bores; Boley \& Durisen 2006}) \acb{and the efficient 
radiative cooling of the disk's upper layers,} even though the midplane is optically 
thick and essentially adiabatic.}

\subsection{\wlad{{\sc BoxzyHydro} simulations}}
\label{sect:BoxzyHydro}

\begin{figure*}
  \begin{center}
    \resizebox{.33\textwidth}{!}{\includegraphics{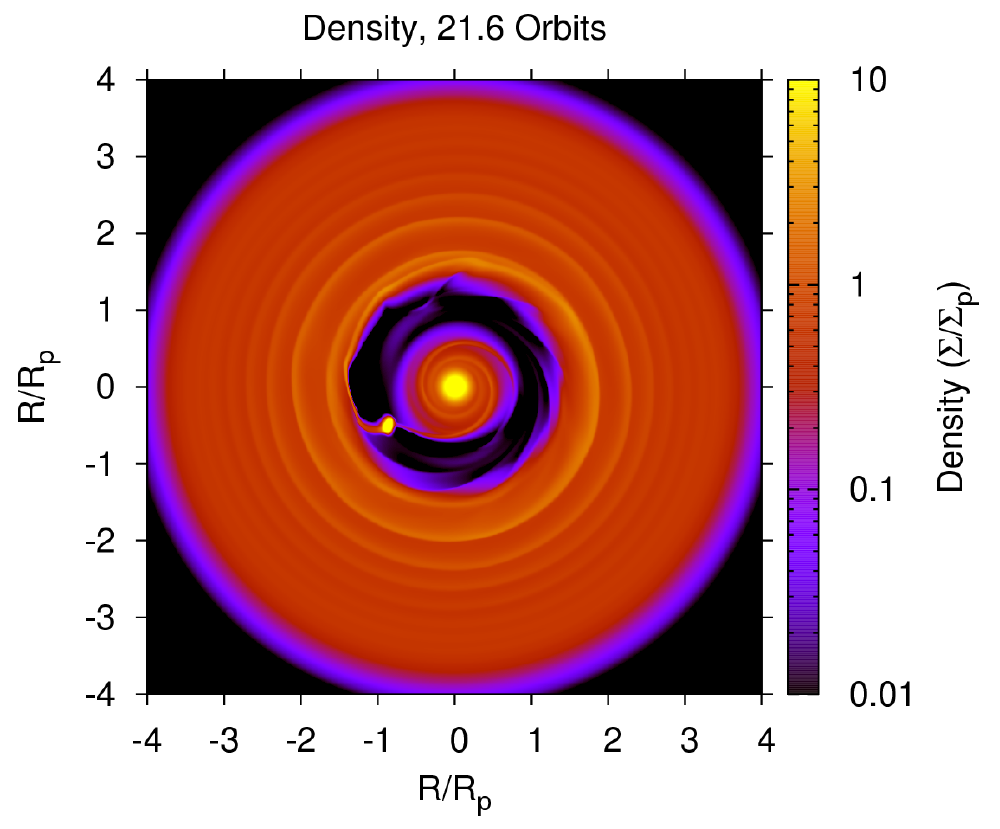}}
    \resizebox{.33\textwidth}{!}{\includegraphics{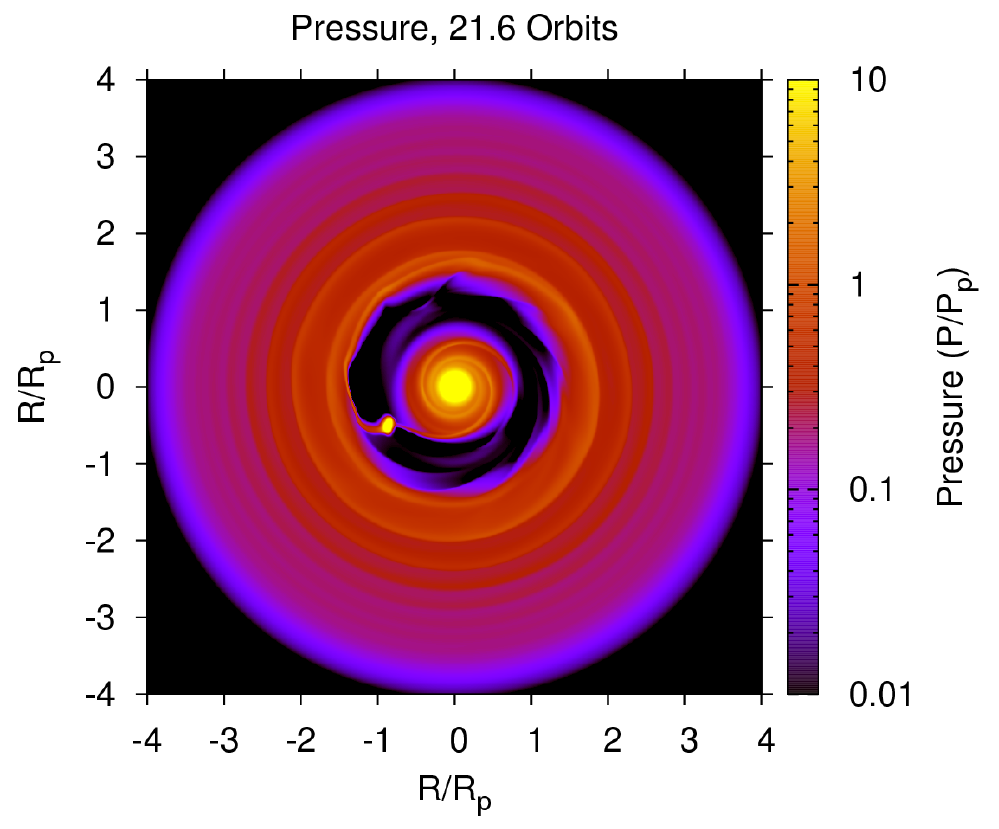}}
    \resizebox{.33\textwidth}{!}{\includegraphics{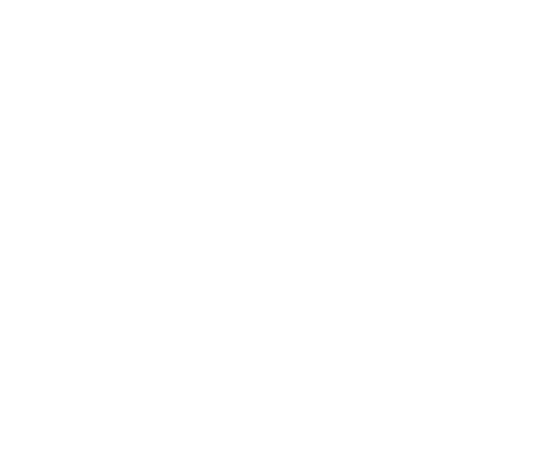}}
    \resizebox{.33\textwidth}{!}{\includegraphics{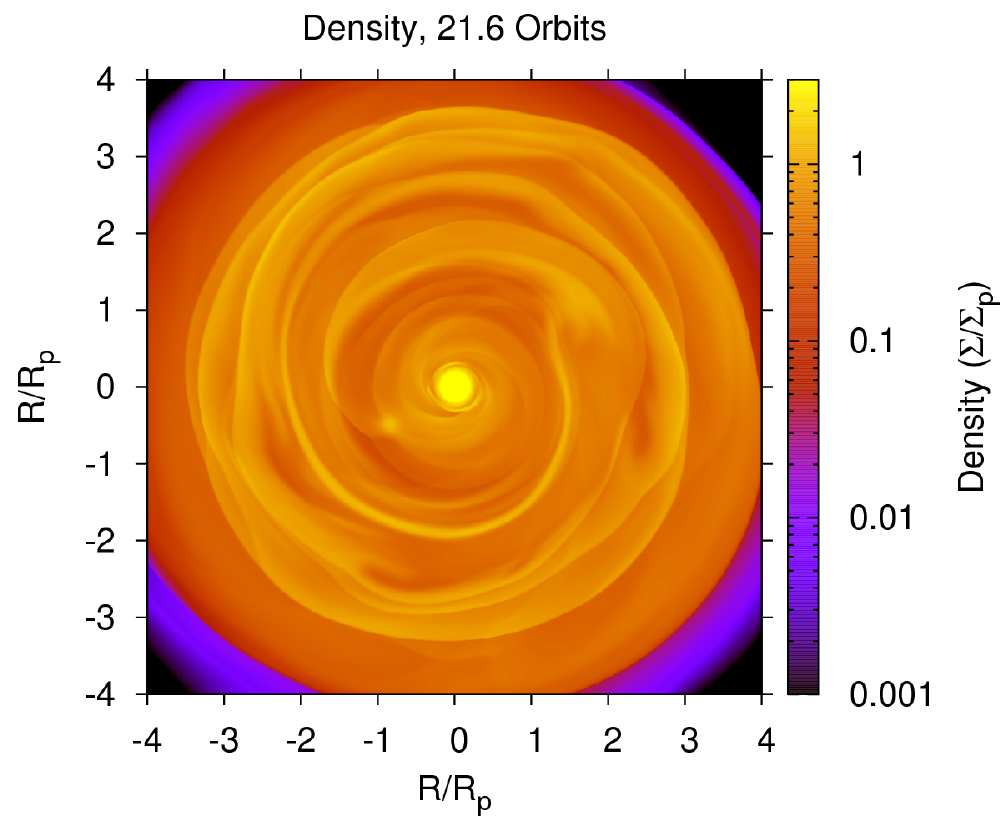}}
    \resizebox{.33\textwidth}{!}{\includegraphics{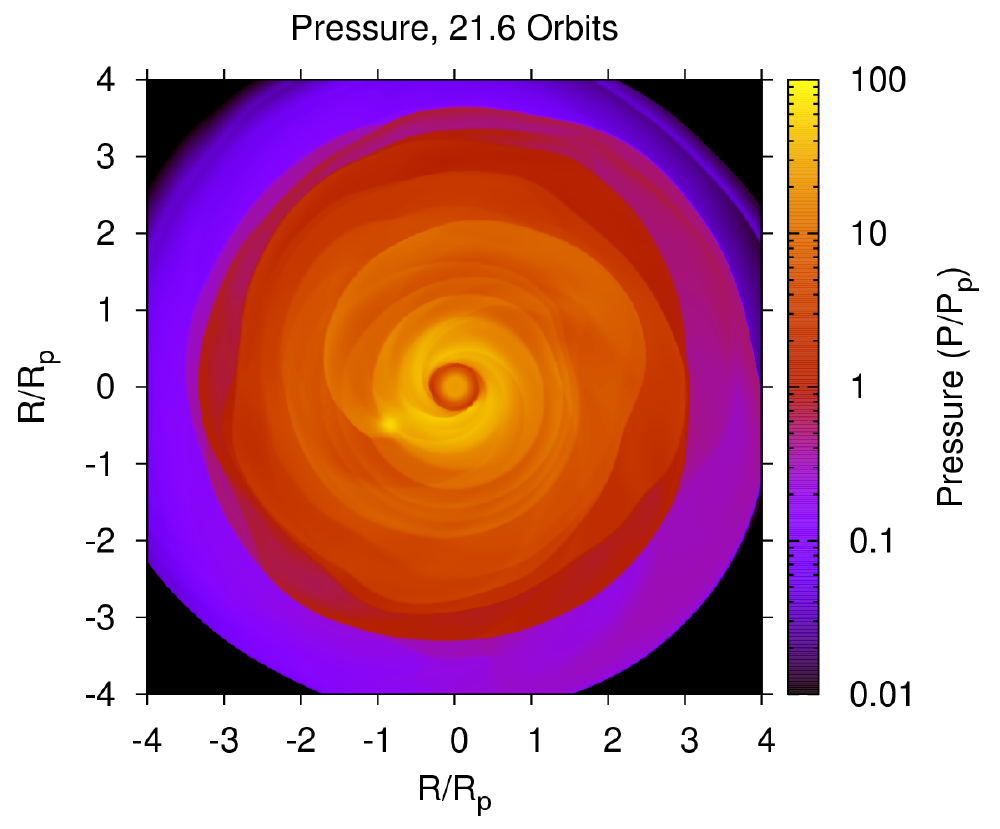}}
    \resizebox{.33\textwidth}{!}{\includegraphics{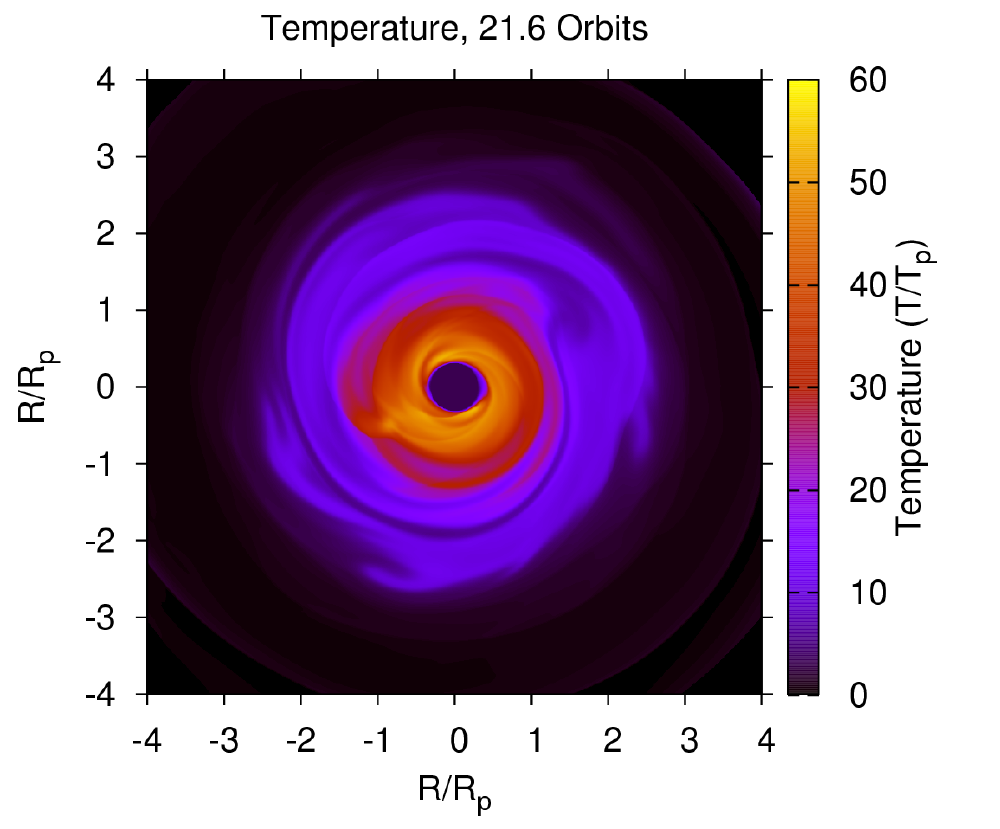}}
  \end{center}
  \caption[]{\wlad{{\sc BoxzyHydro} simulations with a 10$\Mjup$ planet.
  \acb{Surface density, temperature, and pressure are all normalized to the initial values at the planet's semi-major axis.}
     In the upper panels we 
    show an isothermal control run. A clear spiral structure is seen, along with edge waves 
    by the outer gap wells, evidencing the initial stages of vortex formation. In the lower panels 
    we show an adiabatic run, \acb{in which} these features are gone. The shock-driven buoyant instability 
    is reproduced with an independent code. \acb{ {\sc BoxzyHydro} solves for the equations of hydrodynamics in conservative form and uses an approximate Riemann solver for determining fluxes at cell interfaces.  It does not require artificial viscosity terms in the energy equation for shock capturing.}  This rules out the possibility 
    that the instability is effected by Pencil's numerical shock capturing scheme, evidencing a 
    physical origin \acb{for the adiabatic case}.}}
  \label{fig:aaronfigs}
\end{figure*}

\wlad{\citet{muellerkley2013} perform 2D simulations of disks containing protoplanets 
up to 16 MJup using the FARGO code \citep{masset2000,baruteau2008}, 
using isothermal and radiative viscous accretion disk models.
Their results do not show the strong temperature increase seen here.
We first suspected that the difference lie\mordecai{s} in the applied cooling laws: 
while \citet{muellerkley2013} use a cooling function that depends 
on $\rho$ and $T$\mordecai{,} we apply a fixed\mordecai{,} radially dependent cooling. To check this, 
we performed a calculation defining a dynamical cooling time depending on 
$\rho$ and $T$ as well. The results were consistent with the fixed\mordecai{,} radially 
dependent cooling time. Unfortunately, \citet{muellerkley2013} do not
\mordecai{specify whether their code included} artificial viscosity
\mordecai{in order to properly treat} shock heating\mordecai{.  If they did,
it} would enter 
in the\mordecai{ir} viscous energy dissipation term
$Q_\mathrm{visc}$\mordecai{, but its form was not specified}. A meaningful 
comparison between our work and theirs is therefore unfeasible.}

\acb{To \mordecai{resolve this disagreement, and} confirm the occurrence
  of buoyancy-induced turbulence in a  
2D, adiabatic disk with a massive protoplanet, we reproduce run F 
(10$\Mjup$, adiabatic) using an independent code, {\sc BoxzyHydro} \citep{boleyetal2013}. 
{\sc BoxzyHydro} solves the equations of hydrodynamics in conservative form on a Cartesian grid.
An approximate Riemann solver is used, along with standard flux
limiting, \mordecai{to describe} the fluid states at cell interfaces \citep[e.g.,][]{toro2009}.
This allows the code to capture shocks without including explicit artificial viscosity terms in the gas equations. }

\acb{Using {\sc BoxzyHydro}, we model the disk within the radial range $0.4<r<4$ 
(\mordecai{where the planet's semi-major axis} $r_0=1$, as before). The
initial \acb{setup uses} the same density and  
temperature profiles, differing only in the details of \mordecai{the} boundary 
conditions and \mordecai{the} smoothing of the planet gravitational potential. As the 
code is Cartesian, there is no strict inner boundary for the grid. The region 
interior to $r=0.3$ ($\approx$1.6 AU for $r_0=5.2$ AU) is locally isothermal
to avoid intense heating close to the star. The planet potential is piece-wise 
smooth\mordecai{ed} with a cubic spline inside the Hill radius, while keeping the Newtonian 
profile outside. \mordecai{In comparison, in} the Pencil simulations, the
Plummer smoothing \mordecai{used there} (\eq{eq:potential}) 
results in a shallower potential well immediately outward of the Hill radius, therefore 
slightly underestimating the strength of the shocks.}

\wlad{We run the {\sc Boxzyhydro} simulation for 21 orbits. The result is shown 
in \fig{fig:aaronfigs}. The upper panels show an isothermal control run, 
\mordecai{while} the lower panels \mordecai{show} the adiabatic simulations. The isothermal 
run shows well-behaved spiral structure, and edge waves in the outer region of the 
gap, \mordecai{produced by} the growing phase of the \mordecai{RWI. In} the
adiabatic run these features  
are gone, and \mordecai{instead} the code reproduces the main features
seen in Pencil's run F at the same time.  
At this point, turbulence has already begun to proliferate
\mordecai{through} the disk, diminishing the gap (which in run  
F disappears completely at $t\sim50$ orbits)\mordecai{,} and suppressing the 
formation of both inner and outer vortices. \mordecai{These results} rule out the 
possibility that the observed turbulence is an artifact of the shock 
capturing scheme used in the Pencil Code. The confirmation by an independent code 
strengthens the case that the result is \mordecai{not a numerical artifact}}. 

\section{Discussion and Conclusions}
\label{sect:conclusions}

\acb{Using} \mordecai{2D}\acb{,} global \acb{hydrodynamics} simulations of disks with embedded
massive planets, we have shown \wlad{that \acb{shocks generated by}
  high-mass planets} \mordecai{can}
\wlad{significantly \mordecai{raise} the \acb{\mordecai{temperature of gas in} disks}. \acb{While any give\mordecai{n} shock is typically} weak, with Mach numbers of 
order unity for 5$\Mjup$ \acb{and up to} 4--5 for 10$\Mjup$\acb{, t}he temperature increase is due 
to \acb{the cumulative effects of} shocks, as orbiting gas repeatedly
meets the spiral wake (\mordecai{which remains} stationary in the 
reference frame of the planet). If the gas can cool \mordecai{within} a
synodic period, the energy 
radiate\mordecai{s} away between shocks and the situation
resemble\mordecai{s} the isothermal case. If, on the other  
hand, the cooling time \mordecai{exceeds} the synodic period, the mean temperature increase\mordecai{s} in time. 
\mordecai{We then} 
see buoyant instabilities \mordecai{driven by} both the inner and outer spiral wakes,
creating sustained, large-scale turbulence that significantly alters the global structure of the disk.}

\acb{These results are} of particular significance to the frequent interpretation \acb{that} spiral \acb{morphologies} \wlad{\acb{are} signposts of embedded protoplanets in transition disks. \acb{Fits to observed spiral structures are often} attempted 
(e.g.\acb{,} Muto et al. 2012) \acb{by} making use of \acb{linear spiral density wave theory, which depends on 
the temperature gradient in the disk, as well as the actual disk temperatures (disk aspect ratio)}. \acb{However, because this theory is linear, it relies} on assumptions of low planetary mass and local isothermality. In the past few years, 
some spiral features \acb{that} have been observed \mordecai{that}
\acb{present} problems \acb{in interpreting them as having a planetary
  origin, if} linear theory is employed. \acb{For example,} \citet{juhaszetal2014} \mordecai{found}
that the observed spirals in scattered light would require \mordecai{an
  increase by a factor of either} $>3.5$ in surface 
density \mordecai{or $\approx$0.2 in pressure scale height} above the background disk to generate them. \mordecai{T}hey 
favor the increase in pressure scale height \acb{ to} reproduce the observations. This is \acb{consistent with} the observed spirals \acb{showing} in general 
an opening angle wider than expected from the background temperature. Spiral fitting 
to the disk MWC\,758 \citep{benistyetal2015} at \mordecai{radii} $r \approx $80--150\,AU
needs \acb{a} large \mordecai{aspect ratio}, corresponding
to \mordecai{to a disk temperature of 300~K} at 55~AU, while the 
surrounding gas 
is \acb{only} at $\approx$50K. \acb{Finally, r}ecent 
observations of HD\,100546 \citep{currieetal2014} show a planetary candidate without an associated spiral, 
and a feature that resembles a spiral arm, but at 90$^\circ$ away from the candidate planet. Moreover, the 
spiral feature shows little polarization, implying thermal emission at roughly 1000\,K. These features are 
a challenge to linear spiral density theory but \acb{can} easily \acb{be} explained in light of our model, \acb{which} predicts a 
significantly different qualitative behavior for \mordecai{spiral structures
  generated by} high-mass planets in radiatively inefficient disks.} 

\wlad{In the following paper in this
series, we will \mordecai{describe} three-dimensional simulations to \mordecai{further} explore the role of
shock heating from \mordecai{the wakes of} massive planets in the structure and evolution of the disk. 
We caution again that in three dimensions the results will likely differ. Shocks 
should be weaker for \acb{the} same planetary mass, given the possibility of vertical expansion. 
It is possible that the turbulence seen in this work would be seen in another region 
of parameter space; for instance, it may require a higher \acb{protoplanet mass} 
to achieve an effect comparable to that observed in the current work. \acb{Moreover}, 
even though the midplane of the disk is optically thick and close to adiabatic, the 
upper layers \acb{\mordecai{might}} cool efficiently. 
In this case, \acb{the diffusion timescale in the vertical direction may control a given disk's evolution.}
We \acb{note, however, that the buoyant instability may share similarities to other shock-driven processes, such as gravitational instability (Durisen et al.~2007)},
where reasonable
agreement has been shown between 2D and three-dimensional simulations
(see Rice et al.~2014 and 
references therein). Evidence for similar qualitative \acb{behavior} is seen in Boley 
\& Durisen (2006), who modeled shocks due to 2.5$\Mjup$ planets, observing shock 
bores as gas accelerates upwards and breaking waves as the gas descends back 
onto the disk, \mordecai{again} generating turbulence\mordecai{, though by a
  different mechanism}.}

We \acb{finally note} that \mordecai{heating and} buoyancy-induced
turbulence may also affect planet migration. Investigating this effect 
\citep[as in][but for high-mass protoplanets]{zhuetal2012} would require simulations to be
carried out over many more orbits, and for the gravitational back-\mordecai{reaction} of the
gas on the planet to be included.

\acknowledgments AR \mordecai{is funded} by NSF AAG grant AST10-09802, and by the Center for
Exoplanets and Habitable Worlds (PSU). WL is funded by the National
Aeronautics and Space Administration (NASA) through the Sagan Fellowship Program
executed by the NASA Exoplanet Science Institute.
\acb{AB is funded, in part, by the Canada Research Chairs program and
  The University of British Columbia.} \mordecai{M-MML is funded, in
  part, by NASA OSS grant NNX14AJ56G.}
 This work was
performed in 
part at the Jet Propulsion Laboratory, under contract with the California 
Institute of Technology (Caltech). The authors acknowledge discussions with Sijme-Jan
Paardekooper, Axel Brandenburg, Dhrubaditya Mitra\wlad{, Thayne Currie, and \mordecai{Wilhelm} Kley}.

\end{document}